\documentclass{ws-rv9x6}
\usepackage{subfigure}   
\usepackage{ws-rv-thm}   
\usepackage{ws-rv-van}   
\usepackage{lineno}
\makeindex

\begin{document}

\chapter[Studies of double parton scattering in ATLAS ]{Studies of double parton scattering in ATLAS}\label{ra_ch1}

\author[E. M. Lobodzinska on behalf of the ATLAS Collaboration]{E. M. Lobodzinska on behalf of the ATLAS Collaboration}

\address{Deutsches Elektronen Synchrotron Laboratory,\\
Notkestr. 85 22607 Hamburg, Germany \\
ewelina@mail.desy.de}

\begin{abstract}
In this contribution, Double Parton Scattering  processes observed with the ATLAS detector at LHC are discussed. Results of five analyses are  presented:
production of  W boson in association with 2 jets, production of $J/\psi$ meson in association with  W boson, $J/\psi$ production with  Z boson, $J/\psi$ pair production and four jet events. 
\end{abstract}
\body


\section{Introduction}\label{ra_sec1}
Interactions involving more than one pair of incident partons in the same collision have been discussed since the introduction of the parton model to the description of particle production in collisions with hadronic initial states \cite{theory1,theory2}. Double Parton Scattering (DPS) i.e. the interaction of two quark pairs is the simplest case of multi-parton interactions. For a process in which a final state A + B is produced at a centre-of-mass energy $\sqrt{s}$ the simplified formalism of Ref.~\refcite{theory4} gives the equation for DPS cross section:
\begin{equation}
d\sigma^{DPS}_{A+B}(s) = \frac{1}{1+\delta_{AB}}\frac{d\sigma_A(s)d\sigma_B(s)}{\sigma_{e\!f\!f}(s)},
\label{sigma}
\end{equation}
where $\delta_{AB}$ is the Kronecker delta building a symmetry factor, which for the identical final state 
equals 1 and 0 otherwise.
$\sigma_{A}$ and $\sigma_{B}$ are the A and B process cross sections.  $\sigma_{e\!f\!f}$ is a purely phenomenological parameter related to degree of overlap between the interacting hadrons in the plane perpendicular to the direction of motion, determining the overall size of the DPS cross section. $\sigma_{e\!f\!f}$ is assumed to be independent of process, cut and centre-of-mass energy.
Solving  equation \ref{sigma} for $\sigma_{e\!f\!f}$ and expressing the DPS cross section as a total cross section for A+B final state production $\sigma_{A,B}^{tot}$ multiplied by a fraction of DPS events $f_{D\!P\!S}$ gives:
\begin{equation}
\sigma_{e\!f\!f} = \frac{1}{1+\delta_{AB}}\frac{d\sigma_A(s)d\sigma_B(s)}{f_{D\!P\!S}\cdot \sigma_{A,B}^{tot}(s)}.
\label{sigmaeff}
\end{equation}
$\sigma_{A,B}^{tot}$ is understood as a sum of the cross section for single parton scattering (SPS) producing A+B final state and the cross section for double parton interaction (DPS) resulting in the same final state. In the model we use to describe DPS, the processes A and B are assumed to be independent.
 
\section{DPS in W($\rightarrow l\nu$) + 2 jets}
\label{W}
The first process analysed in ATLAS for the presence of DPS was the production of $W$ bosons in association with two jets at a center-of-mass-energy $\sqrt s $ = 7 TeV \cite{dps1}. The data used for this analysis correspond to the integrated luminosity of 36 pb$^{-1}$.

Jets are defined using anti-k$_t$ algorithm with radius parameter R=0.4. They are required to have transverse energy larger than 20 GeV and rapidity $|y| < $ 2.8. 
The signature of DPS are di-jets going back-to-back in the azimuthal plane, because the kinematics of W boson and the di-jet systems are decorelated. This suggests, that  the difference in azimuthal angle allows to distinguish between single and double parton interactions. However, due to to the distortions of this variable by various systematic effects, the balance in the transverse momenta of the jets is used instead. 
It is defined as:
$\Delta_{jets} = |\vec{p}_T^1 + \vec{p}_T^2|,$
where $\vec{p}_T^1$ and $ \vec{p}_T^2 $ are the transverse momentum vectors of the two leading jets.
To avoid the dependence on the jet energy scale, the transverse momentum of di-jet system is normalised to the sum of the individual transverse momenta:
$\Delta^n_{jets} = \frac{|\vec{p}_T^1 + \vec{p}_T^2|}{|\vec{p}_T^1| + |\vec{p}_T^2|}$.
To find $\sigma_{e\!f\!f}$ first the $f_{D\!P\!S}$ factor is extracted. It is done with help of two templates:
\begin{itemize}
\item Template A - W and di-jets originate from the hardest scatter. It is simulated using Alpgen+Herwig+Jimmy \cite{aahj,ahhj,ahjj} Monte Carlo event generators.
\item Template B - DPS events - created using di-jet data sample i.e. data events recorded with exactly two jets.
\end{itemize}
\begin{figure}[ht]
\centerline{
  \subfigure[]
     {\includegraphics[width=2.5in]{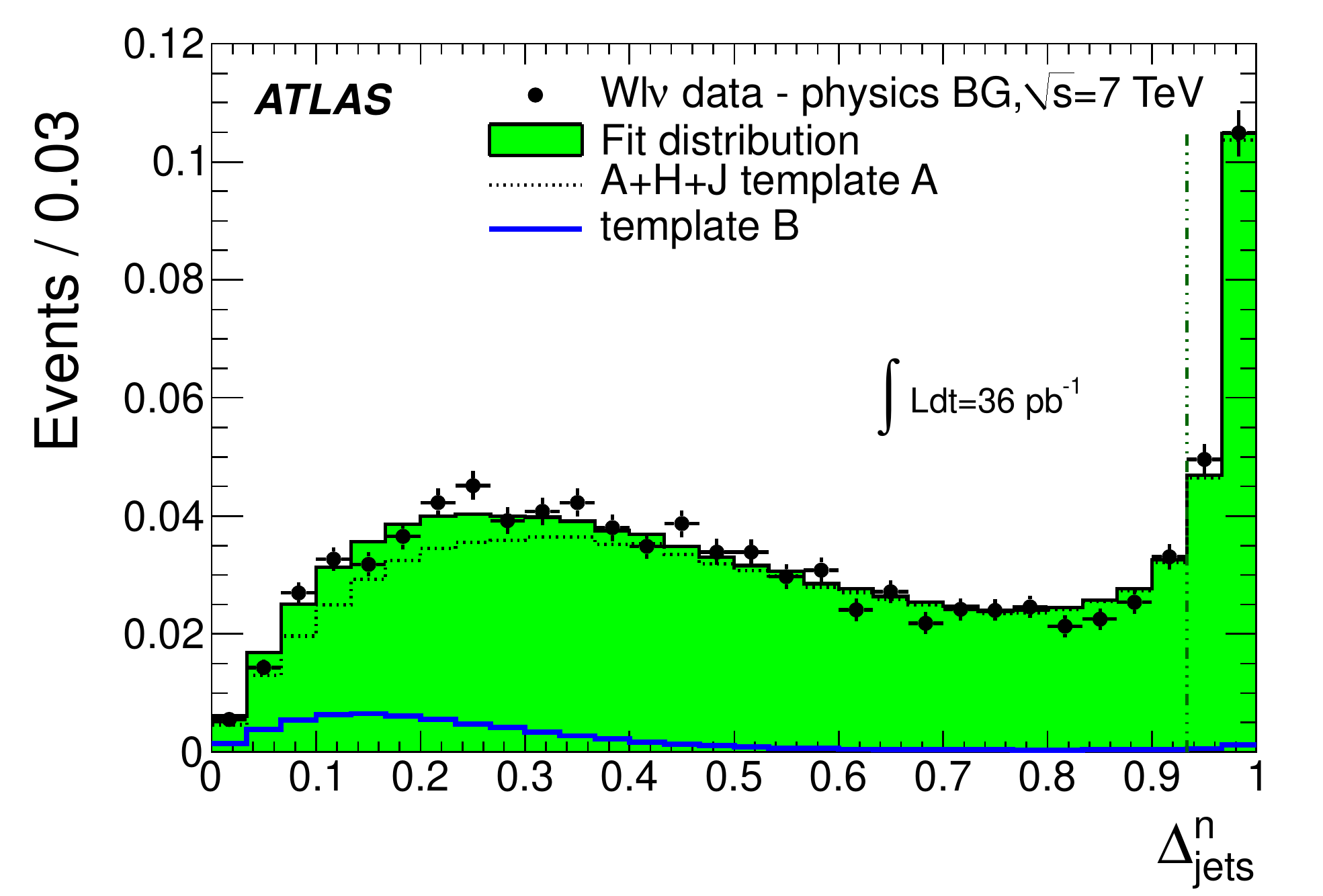}\label{Delta_njets}}
  \hspace*{4pt}
  \subfigure[]
     {\includegraphics[width=2.5in]{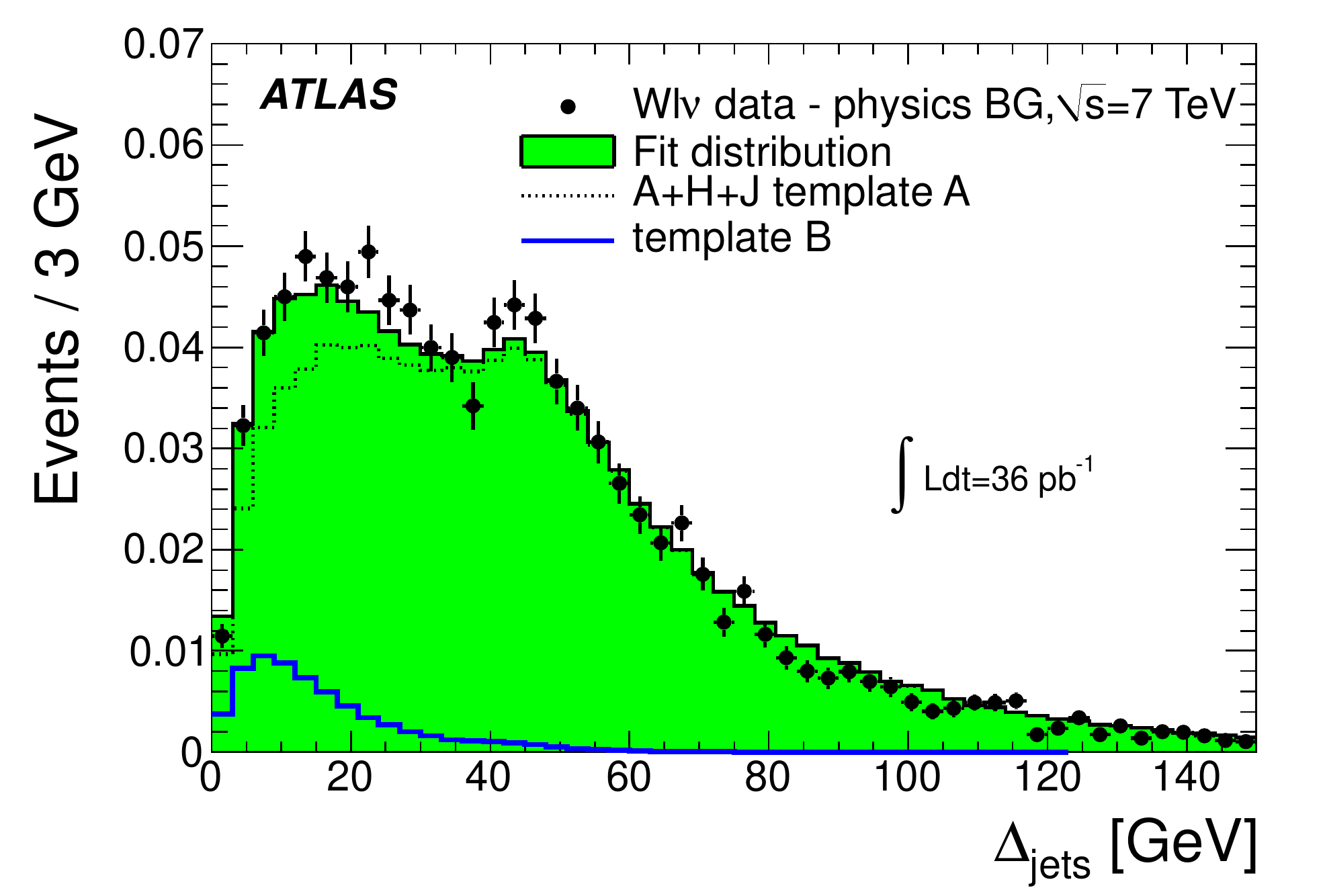}\label{Delta_jets}}
}
\caption{Distribution of $\Delta^n_{jets}$ (a) and $\Delta_{jets}$ (b). The data (shown with dots) and the overall fit ( green histogram) are normalised to unity. Template A (dashed line) and template B (solid line) are normalised to 1-f$_{D\!P\!S}$ and f$_{D\!P\!S}$, respectively. Figure taken from Ref.~\refcite{dps1}} \label{Delta} 
\end{figure}
The data are normalised to unity and fitted by a linear combination of template A and B. From the fit to the distribution
 of $\Delta^n_{jets}$ - see Fig.\ref{Delta_njets} - $F_{D\!P\!S}$ is found to be:
$ f_{D\!P\!S} = 0.08 \pm 0.01 (stat.) \pm 0.02 (syst.).$
As a cross-check the fit is performed also using the $\Delta_{jets}$ distribution - as seen in Fig.\ref{Delta_jets} - and resulting $f_{D\!P\!S}$ is within 13\% of the value received from the $\Delta_{jets}^n$ fit.
The effective cross section is measured to be :
\begin{equation}
\sigma_{e\!f\!f}(7 \textrm{TeV}) = 15 \pm 3(stat) ^{+5}_{-3} (syst.) \textrm{mb}.
\end{equation}
 
\section{DPS in prompt $J/\psi(\rightarrow \mu\mu) + W(\rightarrow \mu\nu)$ }

\begin{figure}[ht] 
\centerline{
\minifigure[Distribution of azimuthal angle between W and $J/\psi$. DPS contribution is shown as a flat template. Hashed area corresponds to the uncertainty on DPS estimation. Figure taken from Ref.~\refcite{dps2}] 
{\includegraphics[width=2.2in]{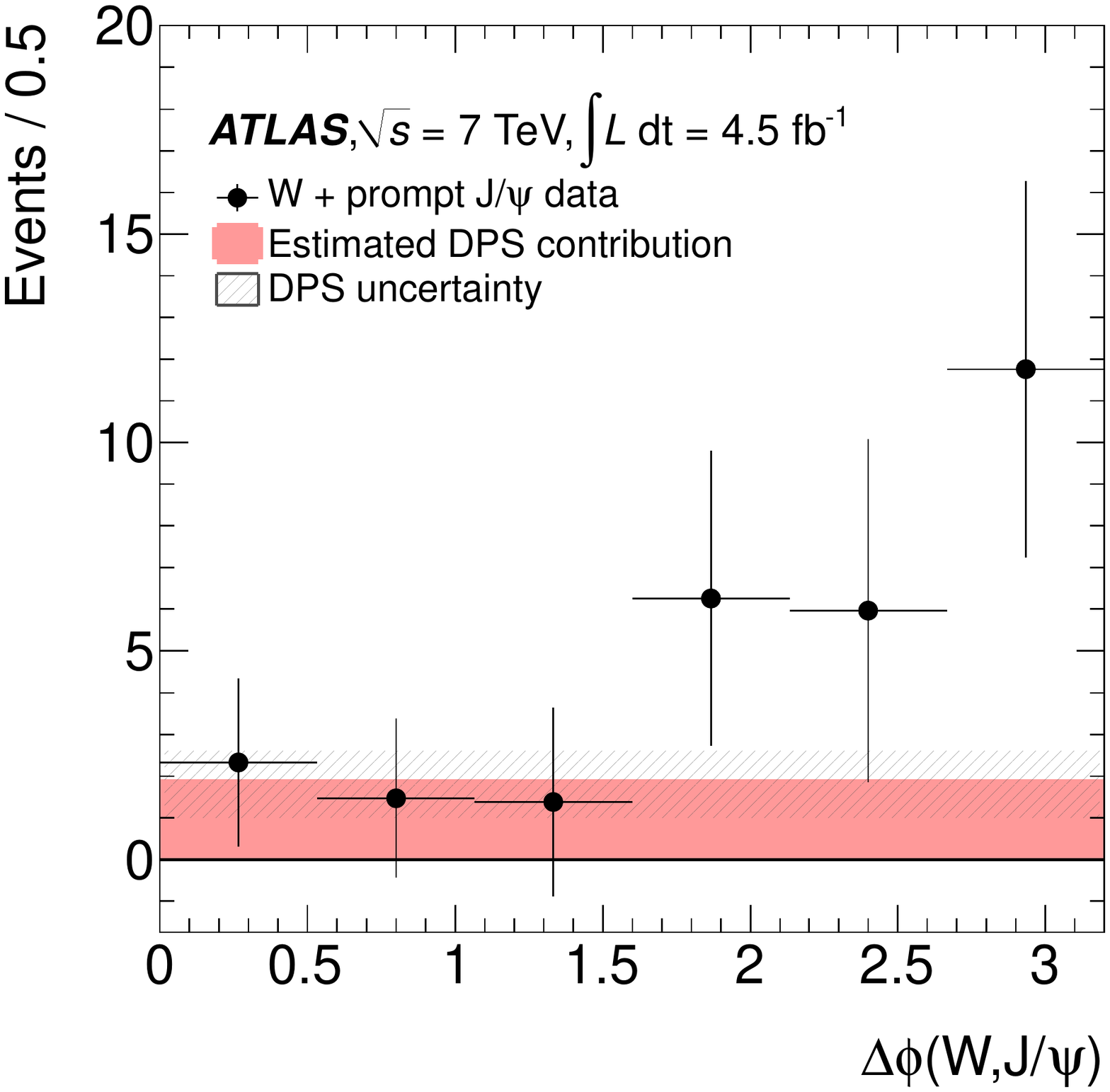}
\label{wjpsi}}
\hspace*{4pt} 
\minifigure[{Distribution of  azimuthal angle between Z and prompt $J/\psi$. DPS contribution is estimated assuming flat template. Hashed region represents the DPS and pile-up uncertainties added in quadrature. Figure taken from Ref.~\refcite{dps3}.} ]
{\includegraphics[width=2in]{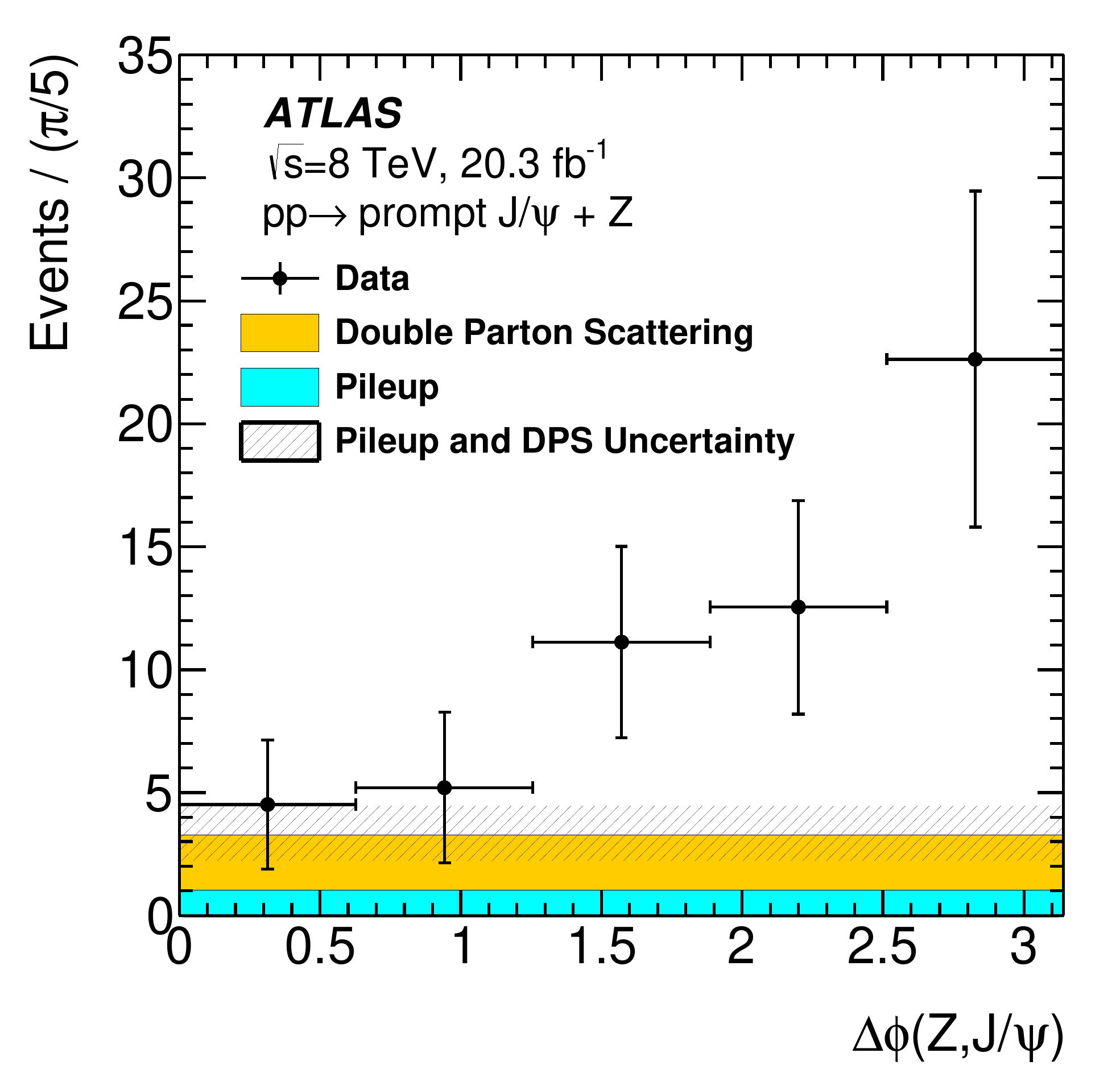}
\label{prompt}}
} 
\end{figure}

The analysis of DPS in prompt $J/\psi$ associated with $W$ production \cite{dps2} uses the 2011 ATLAS dataset with an integrated luminosity = 4.5 fb$^{-1}$ collected at $\sqrt{s}$ = 7 TeV.
Only events containing 3 muons in $|\eta| < $ 2.5 are considered for this analysis. Two oppositely charged muons are required to form $J/\psi$. One muon has to have $p_T > $ 4 GeV, the other one $p_T >$ 3.5 GeV for $|\eta| <$ 1.3 and $p_T >$ 2.5 GeV for $|\eta| >$ 1.3. The invariant mass of the di-muon system is required to be between 2.5 and 3.5 GeV, the transverse momentum of the $J/\psi$ has to satisfy 8.5 $ < p_T^{J/\psi} < $ 30 GeV and the rapidity $ |y_{J/\psi}| < $2.1. 
The third muon is treated as the $W$ boson decay candidate. It has to have $p_T > $ 25 GeV. A momentum imbalance $E_T^{miss}$ larger than 20 GeV and the W boson candidate transverse mass larger than 40 GeV are required.
After events selection and background subtraction a total yield of W+prompt $J/\psi$ events is 29.2$^{+7.5}_{-6.5}$ events.
To calculate the DPS contribution the $\sigma_{e\!f\!f}$ obtained in the analysis of W+di-jet events  and the cross section for prompt $J/\psi$ production $\sigma_{J/\psi}$ as measured in Ref.~\refcite{jpsi_xsec} were used, according to the formula:
\begin{equation}
P_{J/\psi|W} = \frac{\sigma_{J/\psi}}{\sigma_{e\!f\!f}},
\label{formula}
 \end{equation} where $P_{J/\psi|W}$ is the probability of $J/\psi$ production in addition to W already produced in the hard scatter. The DPS contribution is found to be 38$^{+22}_{-20}$\%, which corresponds to  10.8 $\pm$ 4.2 events. Since in DPS the production of W is assumed to be completely independent from the $J/\psi$ production a uniform distribution in azimuthal angle between $J/\psi$ and W is expected. On the contrary, in case of single parton interaction a strong azimuthal correlation  between $J/\psi$ and W is expected. Fig.\ref{wjpsi} shows the distribution of the difference in azimuthal angle $\Delta\phi$ between $J/\psi$ and W.
 \section{DPS in prompt and non-prompt $J/\psi(\rightarrow \mu\mu) + Z(\rightarrow ll)$ }
\label{Z}
A similar analysis was performed for prompt and non-prompt $J/\psi$ production associated with a Z boson \cite{dps3}. ATLAS data collected in 2012 at $\sqrt{s} = $ 8 TeV with an integrated luminosity of 20.3 fb$^{-1}$ were used.
$J/\psi$ selection is the same as in the analysis above, and for Z boson identification two oppositely charged leptons of $p_T > $ 15 GeV and invariant mass of the di-lepton system satisfying 81 $ > m_{ll} > $ 101 GeV are required.
In a sample of events containing a Z boson candidate 56 $\pm$ 10 contained in addition a prompt $J/\psi$ and 
95$\pm$ 12 a non-prompt one.
From  formula \ref{formula} 
the DPS contribution   
was measured to be 29 $\pm$ 9 \% for Z+prompt $J/\psi$ and 8 $\pm$ 2 \% for Z+non-prompt $J/\psi$ events. 
Fig.\ref{prompt} shows the distribution of the difference in azimuthal angle between the Z boson and the prompt  $J/\psi$. A uniform distribution of DPS events is assumed.  The small $\Delta \phi$ region is sensitive to DPS contribution, so 
conservatively assuming that all events in the region of $\Delta\phi < \pi$/5 are due to DPS a lower limit $\sigma_{e\!f\!f} >$ 5.3 mb (3.7 mb) at 68\%(95\%) confidence level is found for Z+prompt $J/\psi$ events.

 
\section{DPS in prompt $J/\psi(\rightarrow \mu \mu)$ pair production}
\begin{figure}[ht]
\centerline{
  \subfigure[]
     {\includegraphics[width=2in]{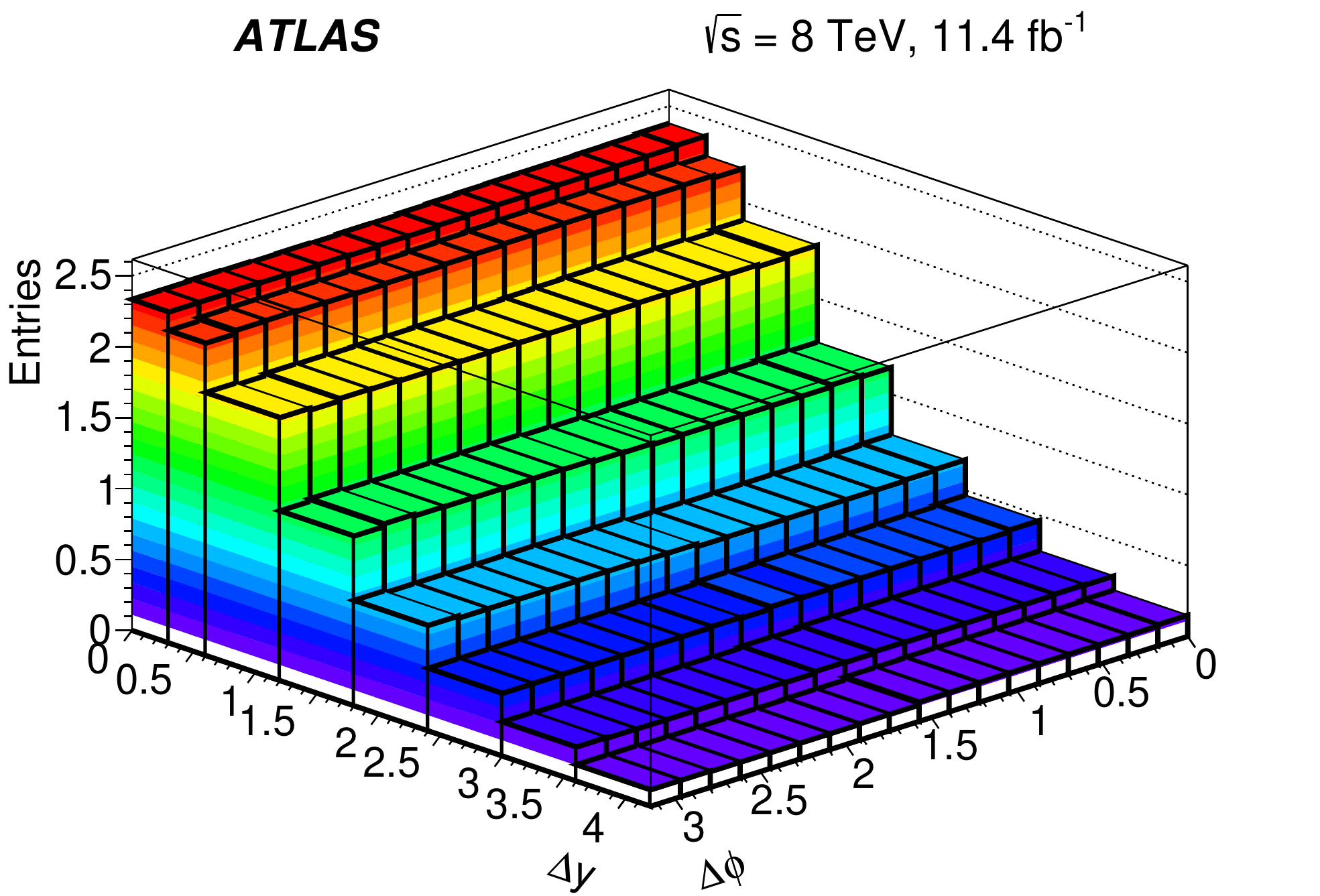}\label{temp1}}
  \hspace*{4pt}
  \subfigure[]
     {\includegraphics[width=2in]{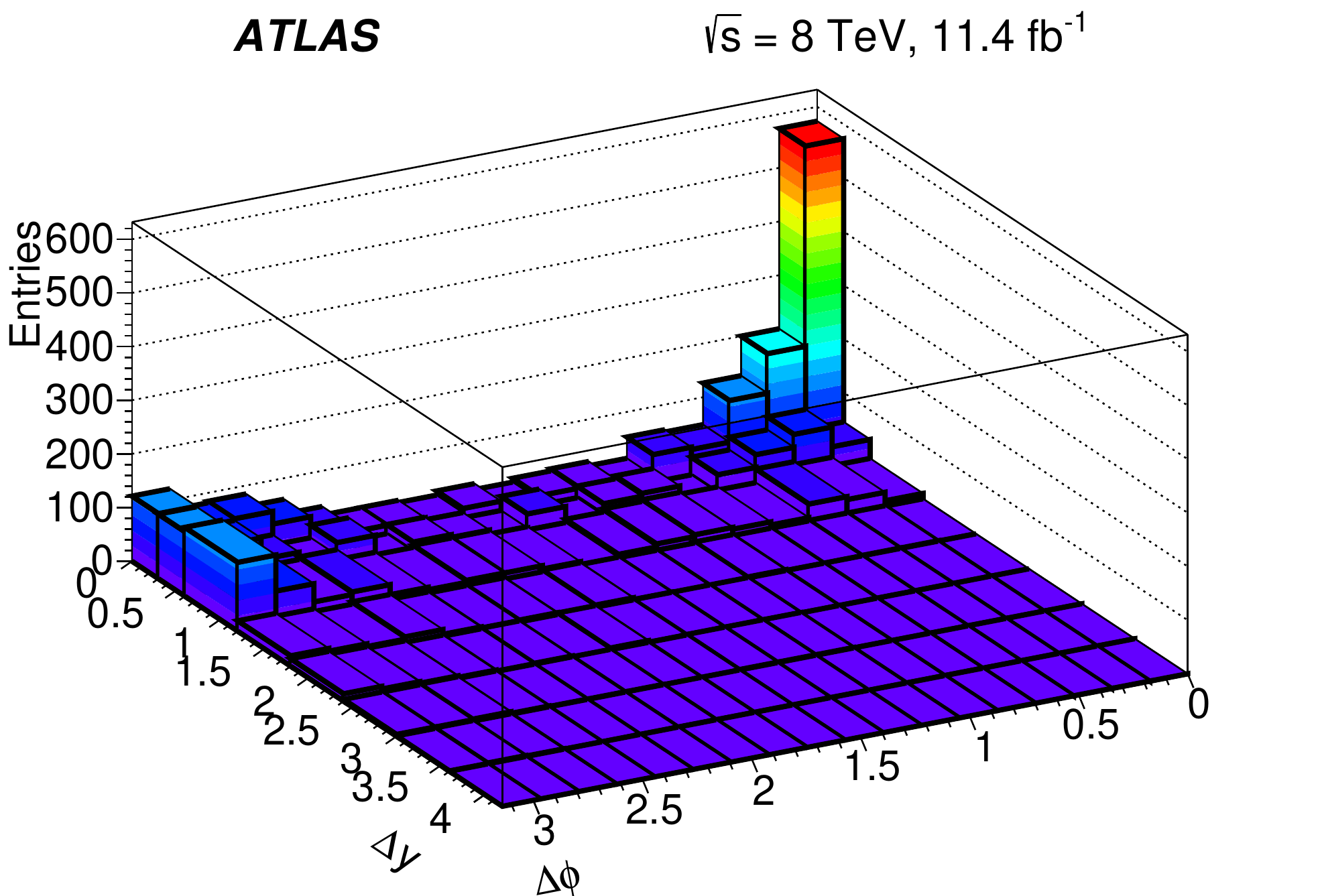}\label{temp2}}
}
\caption{2-dimensional data driven templates of $\Delta y$ against $\Delta\phi$ for (a) DPS and (b) SPS production of  di-$J/\psi$ final state. Figure taken from Ref.~\refcite{dps4}.}\label{ra_fig2}
\end{figure}

\begin{figure}[ht]
\centerline{
  \subfigure[]
     {\includegraphics[width=2in]{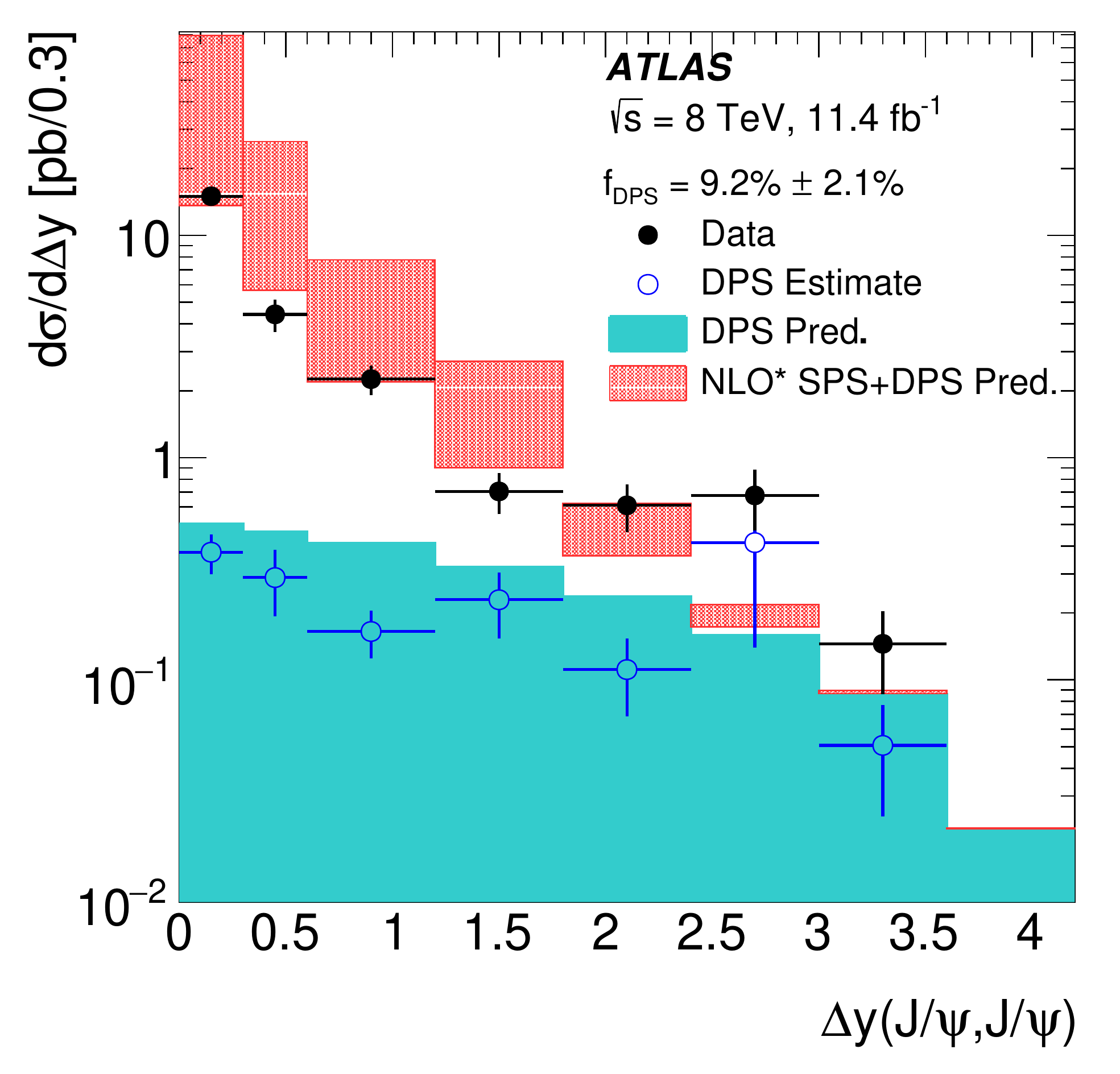}\label{diffxsec_y}}
  \hspace*{4pt}
  \subfigure[]
     {\includegraphics[width=2in]{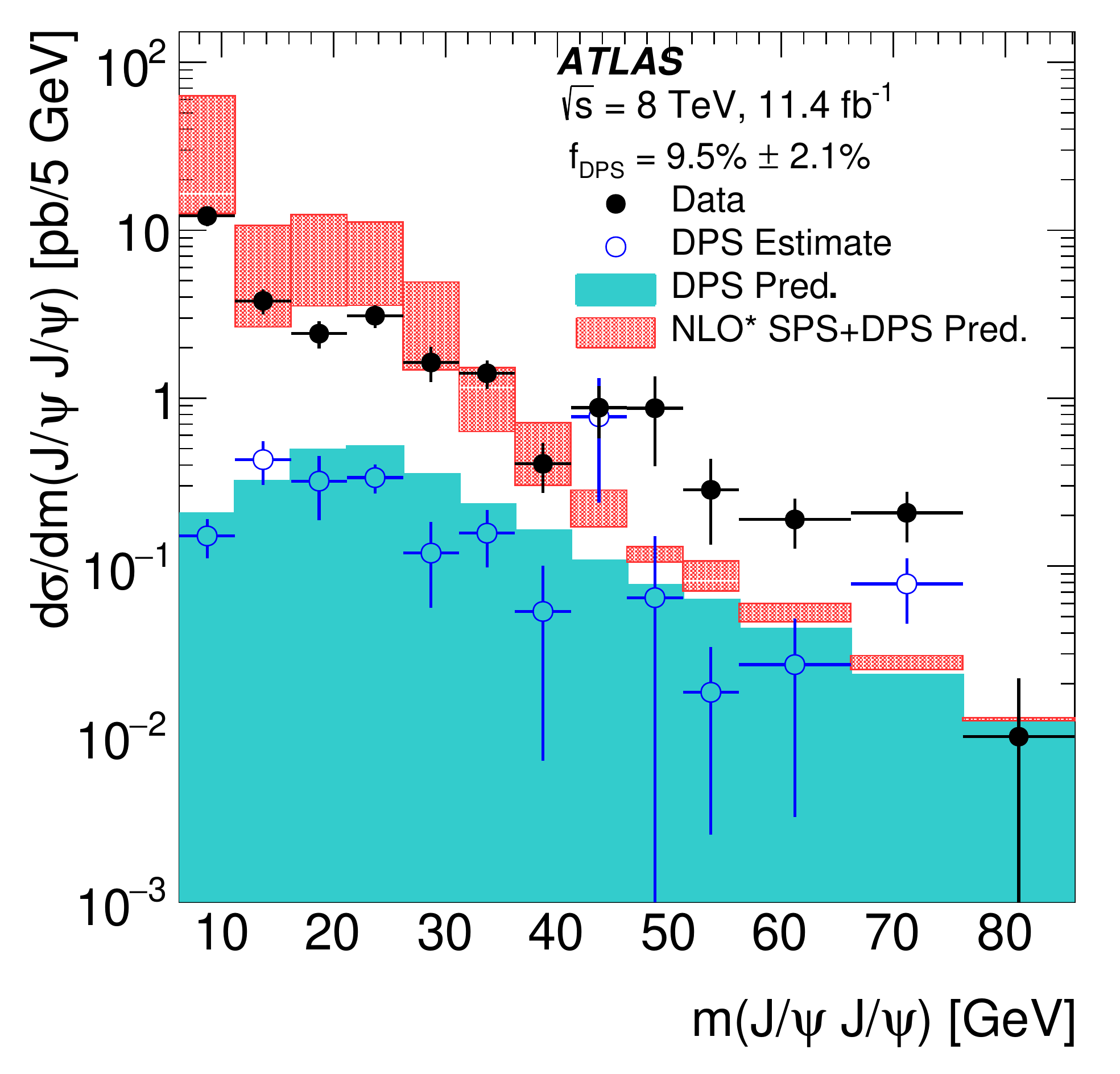}\label{diffxsec_m}}
}
\caption{The DPS and total differential cross sections as a function of (a) difference in rapidity between the two $J/\psi$ mesons, 
(b) the invariant mass of the di-$J/\psi$. 
In addition to the data also the DPS \cite{di43} and NLO SPS \cite{di1011} theoretical predictions are shown. Figure taken from Ref.~\refcite{dps4}} \label{diffxsec} 
\end{figure}

The production of two prompt $J/\psi$ mesons \cite{dps4} is studied using ATLAS data collected in 2012 at $\sqrt{s}$ = 8 TeV with an integrated luminosity of 11.4 fb$^{-1}$. Selected are events containing four muons of $p_T >$ 4 GeV and invariant mass in the range 2.5 $< m(\mu\mu) <$ 4.3 GeV. The muons are supposed to be decay products of two $J/\psi$ mesons, each with $p_T >$ 8.5 GeV and rapidity $|y| < $ 2.1. To assure that the mesons are coming from the same proton-proton collision, the distance between the two $J/\psi$ vertices is required to be smaller that 1.2 mm.
In total 1210 events are selected.
In order to extract the DPS contribution, templates for single and double parton interactions are created and fitted to the data. The DPS template is constructed by assuming that the two  $J/\psi$ candidates are produced independently of each other, so  re-sampled $J/\psi$ mesons from two different random events in the di-$J/\psi$ sample are used. The distribution of the absolute value of a difference in azimuthal angle $\Delta \phi$ between the two $J/\psi$ mesons  against the difference in rapidity $\Delta y$ is shown in Fig.\ref{temp1}. The normalisation is found from the data, by assuming that for $\Delta y > $ 1.8 and $\Delta \phi < \pi/2$ DPS dominates, and SPS is negligible. The template for single parton scattering is obtained by subtracting the DPS template from the background subtracted data and is shown in Fig.\ref{temp2}.
From the 2-dimensional data driven DPS and SPS templates the weights of DPS and SPS events are calculated.
These weights are used to show the SPS and DPS contributions to the total differential cross section for di-$J/\psi$ production as a function of different variables as seen in Fig.\ref{diffxsec}.
This figure shows also  the theoretical predictions for DPS expectations from Ref.~\refcite{di43} and NLO SPS predictions from Ref.~\refcite{di1011}. The DPS prediction correctly describes the data-driven DPS distribution, however the prediction of the SPS shows significant disagreement with the data e.g. at large $\Delta y$ and  large invariant masses.

The contribution of DPS events $f_{D\!P\!S}$ is measured from Fig.\ref{diffxsec_y} and yields
$f_{D\!P\!S} =$ 9.2 $\pm$ 2.1 (stat.) $\pm$ 0.5 (syst).
 Using the $f_{D\!P\!S}$, the cross section for $J/\psi$ production and the total cross section for $J/\psi$ pair production the  effective cross section of double parton scattering is  measured to be: 
\begin{equation}
\sigma_{e\!f\!f} = 6.3 \pm 1.6 (stat) \pm 1.0 (syst) \textrm{mb},
\end{equation} 
which is lower than that measured for other final states e.g. sec.\ref{W}. 

\section{DPS in four-jet events}
This analysis uses  inclusive four-jet events produced in $pp$ collisions at $\sqrt{s}=$ 7 TeV. The data sample corresponds to an integrated luminosity of 37.3 pb$^{-1}$. At least one reconstructed primary vertex is required to reject events originating from non-collision background. The jets are identified using anti-kt algorithm with R=0.6. Each jet is required to have transverse momentum $p_T > $ 20 GeV and pseudorapidity $|\eta| < $ 4.4. To insure 100\% trigger efficiency the leading jet was required to have $p_T^1 > $ 42.5 GeV.

To estimate the fraction of DPS events three templates are constructed:
\begin{itemize}
\item single parton scattering (SPS) - all jets originate from the hardest scattering - this template is modeled using Alpgen+Herwig+Jimmy Monte Carlo \cite{aahj,ahhj,ahjj},
\item complete Double Parton Scattering (cDPS) - the secondary scatter produces two of the four leading jets in the event - a template is constructed by overlying two data di-jet events, 
\item semi Double Parton Scattering (sDPS) - three out of four jets originate from the hardest scatter and only one from the secondary scatter - a template is modeled by overlying data events (3-jet and 1-jet).   
\end{itemize}

In case of cDPS the double di-jet production should result in pairwise $p_T$ balanced jets going back-to-back. Additionally, a flat distribution between azimuthal angles  of both interaction planes is expected, as the processes are independent. sDPS resembles SPS topology, however the fourth jet in sDPS is not expected to show any correlation with other jets, while in SPS such a correlation should be visible.
To distinguish between SPS, cDPS and sDPS several differentiating variables are used e.g.:
$\Delta_{ij}^{pT} = \frac{|\vec{p}_T^i + \vec{p}_T^j|}{p_T^i+p_T^j}, \Delta\phi_{ij} = |\phi_i - \phi_j|$ or $\Delta y_{ij}= |y_i-y_j|,$ 
where the index $i$ and $j$ denote the jet number and subscript $ij$ indicates that all possible jet combinations are studied. 
$\phi_{i+j}$ stands for azimuthal angle of the four-vector obtained by the sum of jets $i$ and $j$.
The analysis shows that none of the variables alone will distinguish between all three components, but all are important and should be taken into account. In total 21 variables are constructed and used as an input to train a neural network (NN). The output of the NN consists of three variables, which are interpreted as the probability of an event to be SPS ($\xi_{SPS}$), cDPS ($\xi_{cDPS}$) and sDPS ($\xi_{sDPS}$). 
Each event is plotted as a single point on an equilateral triangle (ternary plot) using the constrain $\xi_{SPS}+\xi_{cDPS}+\xi_{sDPS}$ = 1. 
From the fit parameters the fractions of DPS and SPS events are found. The total $f_{D\!P\!S}$ is $0.092^{+0.005}_{-0.011}(stat.)^{+0.033}_{-0.037}(syst.)$, and the sDPS contribution to the $f_{D\!P\!S}$ is found to be 40\%. 
\begin{figure}[ht]
\centerline{
  \subfigure[]
     {\includegraphics[width=2in]{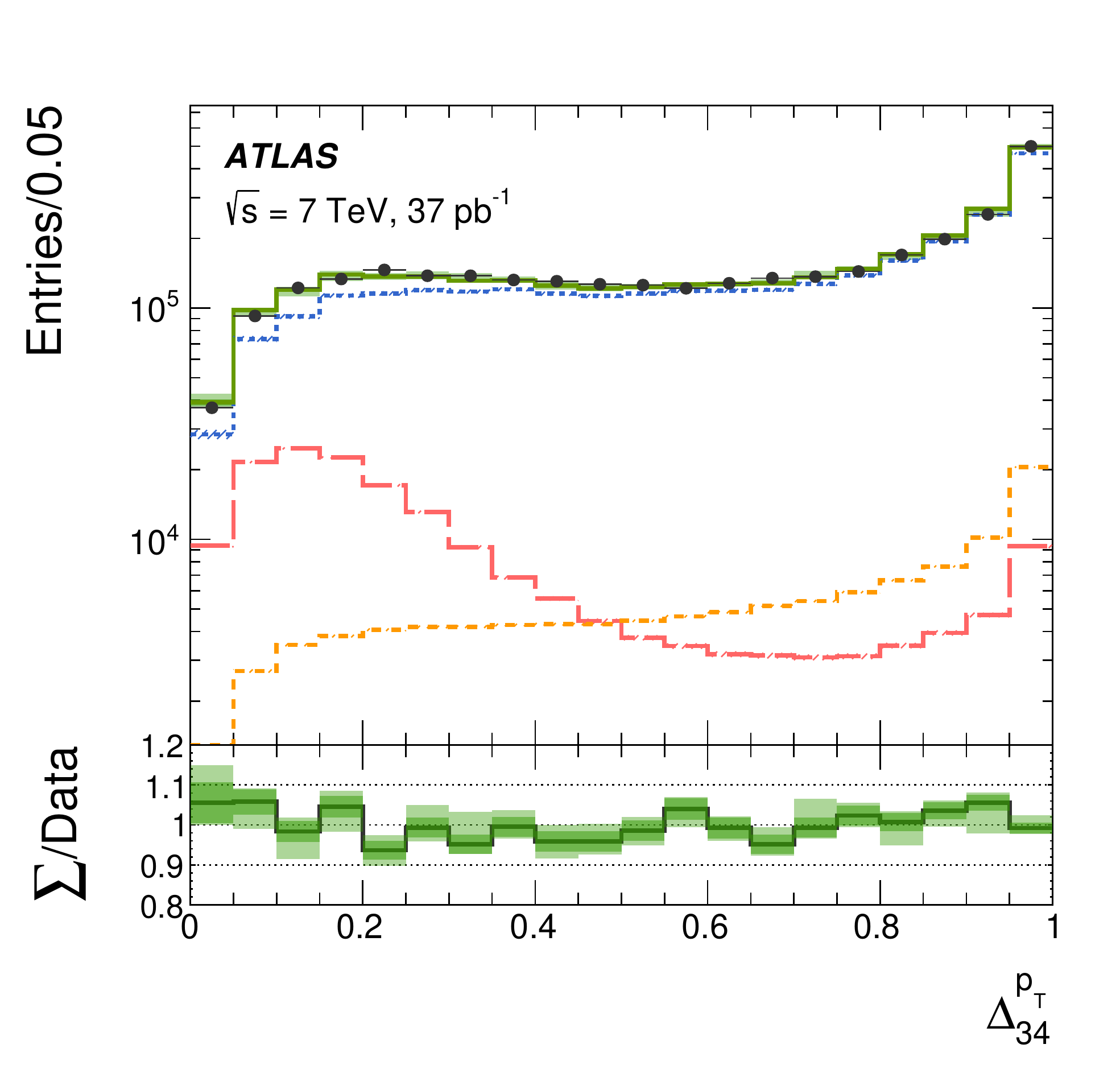}\label{validation_a}}
  \hspace*{4pt}
  \subfigure[]
     {\includegraphics[width=2in]{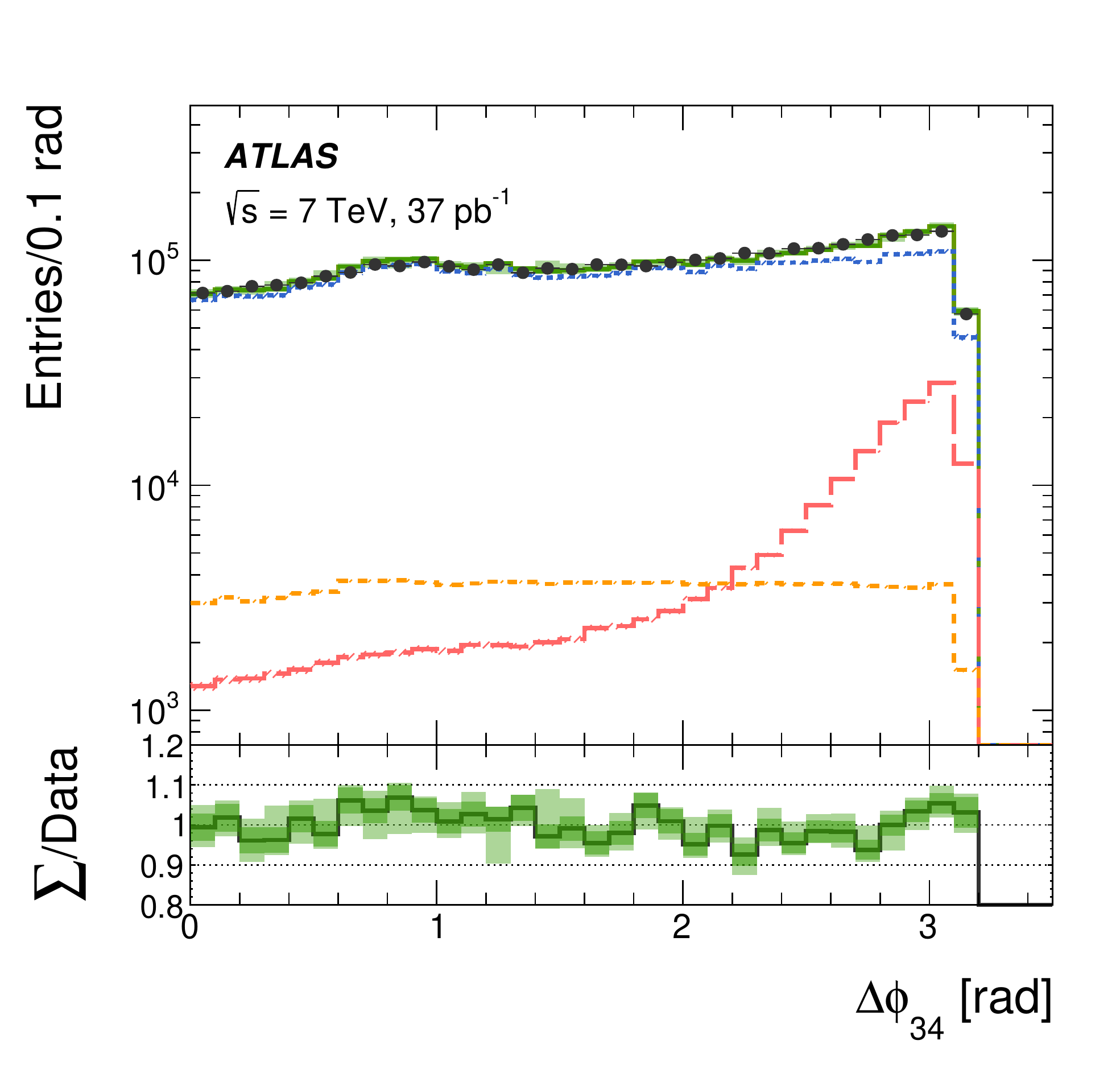}\label{validation_b}}
}
\caption{Comparison between the distributions of the variables (a) $\Delta_{34}^{p_T}$ and (b) $\Delta\phi_{34}$ in 4-jet events in the data and the sum of the SPS, cDPS and sDPS contributions. The sum of the contributions is normalised to the cross section measured in data and the various contributions are normalised to their respective fractions obtained from the fit. Figure taken from Ref.~\refcite{dps5}.}\label{validation}
\end{figure}
 
In order to validate the result the data distributions for different variables are compared to the sum of SPS, cDPS sDPS contributions obtained from the NN. In Fig.\ref{validation} the distributions of  $\Delta^{pT}_{34}$ and $\Delta\phi_{34}$ are shown. A good overall agreement is observed.
Using di-jet and 4-jet cross sections and the $f_{D\!P\!S}$ fraction the $\sigma_{e\!f\!f}$ value was extracted, yielding:
\begin{equation}
\sigma_{e\!f\!f} = 14.9^{+1.2}_{-1.0} (stat.) ^{+5.1}_{-3.8} (syst.) \textrm{mb}.
\end{equation}
This number is consistent with earlier ATLAS measurements described in sections \ref{W} and \ref{Z}.

\section{Summary}
Presented are five measurements related to the DPS analysis in ATLAS. W+2jets, prompt $J/\psi$ pair and 4-jets provide measurements of $\sigma_{e\!f\!f}$. $Z+J/\psi$ estimates a lower limit on $\sigma_{e\!f\!f}$ and $W+J/\psi$ shows sensitivity to DPS.
\begin{figure}
\centerline{\includegraphics[width=6.8cm]{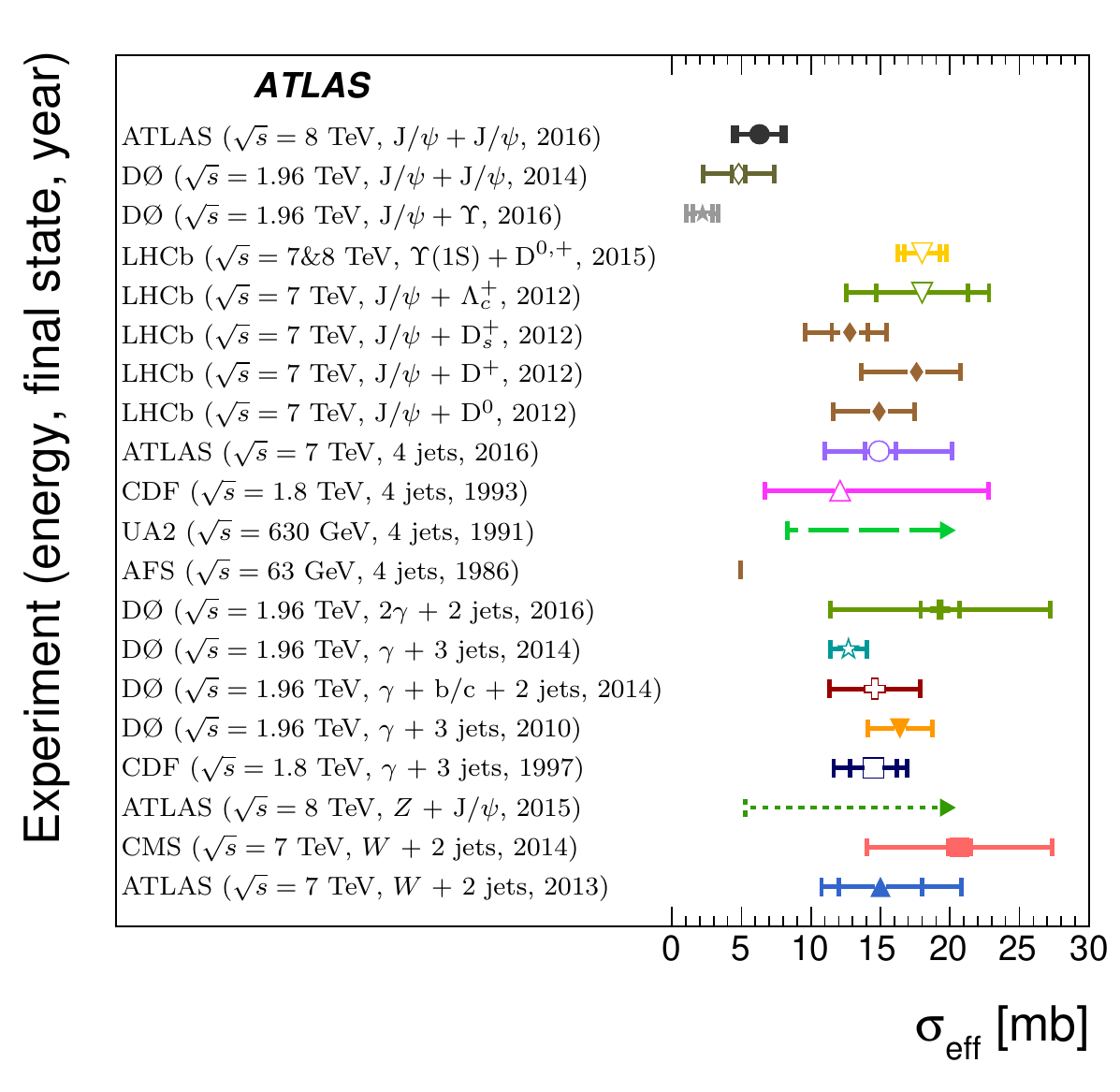}}
\caption{The effective cross section of DPS as measured using various energies and final states by different experiments \cite{afs,ua2,cdf1,cdf2,d01,d02,d03,d04,d05,cms,lhcb1,lhcb2,dps1,dps3,dps4,dps5}. Dashed arrows indicate lower limits. Figure taken from Ref.~\refcite{dps4}} \label{compar}
\end{figure}
In Fig.\ref{compar} the ATLAS results for effective sigma are compared to measurements done at different energies and final states by various experiments. 
Generally all measurements of $\sigma_{e\!f\!f}$ are consistent between each others, only the measurement done with the $J/\psi$ pair tend to have a  lower effective sigma.




\end{document}